\documentclass[english,aps,preprintnumbers,amsmath,amssymb,nofootinbib,twocolumn,10pt]{revtex4-1} 

\usepackage[latin1]{inputenc}
\usepackage{babel}
\usepackage{graphicx}
\usepackage{psfrag}
\usepackage{color}
\usepackage{dsfont}

\usepackage{slashed}

\usepackage{mathbbol}

\newcommand{\psfragnumbers}{
  \psfrag{-}{$-$}
  \psfrag{.}{$.$}
  \psfrag{0}{$0$}
  \psfrag{1}{$1$}
  \psfrag{2}{$2$}
  \psfrag{3}{$3$}
  \psfrag{4}{$4$}
  \psfrag{5}{$5$}
  \psfrag{6}{$6$}
  \psfrag{7}{$7$}
  \psfrag{8}{$8$}
  \psfrag{9}{$9$}}
\newcommand{\Nf}{N}
\newcommand{\Nffour}{N_{\text{f}}}
\newcommand{\Eqref}[1]{Eq.~\eqref{#1}}
\newcommand{\ttm}[1]{\ensuremath{\text{\tiny{$#1$}}}}
\newcommand{\euler}{\mathrm{e}}
\newcommand{\cplx}{\mri}
\newcommand{\arcosh}{\operatorname{arcosh}}

\newcommand{\tr}{\operatorname{tr}}

\newcommand{\STr}{\operatorname{STr}}
\newcommand{\abs}[1]{\ensuremath{\left\vert#1\right\vert}}
\newcommand{\regint}[1]{\int \!\!\! {}_{{}_{{}_{{}_{{}_\text{\small{\ensuremath{#1}}}}}}}}
\newcommand{\mcA}{\ensuremath{\mathcal{A}}}

\newcommand{\mcF}{\ensuremath{\mathcal{F}}}

\newcommand{\mcO}{\ensuremath{\mathcal{O}}}
\newcommand{\mcP}{\ensuremath{\mathcal{P}}}

\newcommand{\mrd}{\ensuremath{\mathrm{d}}}

\newcommand{\mrf}{\ensuremath{\mathrm{f}}}

\newcommand{\mrG}{\ensuremath{\mathrm{G}}}
\newcommand{\mri}{\ensuremath{\mathrm{i}}}
\newcommand{\mrI}{\ensuremath{\mathrm{I}}}

\newcommand{\mrP}{\ensuremath{\mathrm{P}}}
\newcommand{\mrT}{\ensuremath{\mathrm{T}}}

\newcommand{\mfrF}{\ensuremath{\mathfrak{F}}}
\newcommand{\mfrI}{\ensuremath{\mathfrak{I}}}

\newcommand{\rmi}{\ensuremath{(\mathrm{i})}}
\newcommand{\rmii}{\ensuremath{(\mathrm{ii})}}
\newcommand{\rmiii}{\ensuremath{(\mathrm{iii})}}

\makeatother

\begin{document}

\title{Renormalization flow towards gravitational catalysis in the $3d$ Gross-Neveu model}
\author{Holger Gies and Stefan Lippoldt}
\affiliation{Theoretisch-Physikalisches Institut, Friedrich-Schiller-Universit\"at Jena, 
Max-Wien-Platz 1, D-07743 Jena, Germany}                                        

\begin{abstract}
  Catalyzed symmetry breaking arises from a parametric enhancement of
  critical fluctuations independently of the coupling
  strength. Symmetry-breaking fermionic long-range fluctuations
  exhibit such an enhancement on negatively curved spaces, as is known
  from mean-field studies. We study gravitational catalysis from the
  viewpoint of the functional renormalization group using the $3d$
  Gross-Neveu model as a specific example. We observe gravitational
  catalysis towards a phase of broken discrete chiral symmetry both on
  a maximally symmetric (AdS) and on a purely spatially curved
  manifold for constant negative curvature (Lobachevsky
    plane). The resulting picture for gravitational catalysis
  obtained from the renormalization flow is closely related to that of
  magnetic catalysis. As an application, we estimate the curvature
  required for subcritical systems of finite length to acquire a
  gravitionally catalyzed gap.
\end{abstract}

\maketitle

\section{Introduction}\label{sec:intro}

Mass gap generation and (chiral) symmetry breaking in relativistic fermionic
systems can arise from a variety of mechanisms which are often related to
certain couplings or interaction channels becoming dominant. An apparent
counter-example is catalyzed symmetry breaking, first studied in the context
of magnetic catalysis \cite{Gusynin:1994va,Gusynin:1994xp,Gusynin:1995nb},
where mass gap generation is triggered by the presence of a magnetic field
even for arbitrarily small values of the interaction strength. This phenomenon
can be understood in various ways, the essence being that the long-range
fluctuations driving the symmetry-breaking transitions are parametrically
enhanced, see \cite{Shovkovy:2012zn} for a recent review. More concretely, a
magnetic field induces a fermionic fluctuation spectrum with Landau-levels
containing a zero mode that leads to an enhancement of the density of states
in the IR and to an effective dimensional reduction favoring symmetry
breaking.  Magnetic catalysis has found a rich variety of applications both in
particle physics (chiral phases of QCD) \cite{Shushpanov:1997sf,%
  Cohen:2007bt,Zayakin:2008cy,Mizher:2010zb,Gatto:2010qs,Boomsma:2009yk,Bali:2012zg}
and condensed matter physics
\cite{Semenoff:1998bk,Krishana:1999,Khveshchenko:2006,Khveshchenko:2001zza,Leal:2003sg,Herbut:2008ui,Ferrer:2008dy}.

Another simple picture for magnetic catalysis has recently been developed
within the framework of the functional renormalization group in the context of
the $3d$ Gross-Neveu model \cite{Scherer:2012nn}. In line with the fact that
symmetry-breaking phase transitions are often related to fixed points of
renormalization group transformations, also magnetic catalysis can be related
to the behavior of RG fixed points as a function of the magnetic field. This
RG picture has already successfully been applied in the context of QCD
\cite{Fukushima:2012xw}. 

In the present work, we verify the underlying RG picture of catalyzed symmetry
breaking in the context of curved spacetimes. The fact that
symmetry breaking and mass generation in fermionic systems can be influenced by
negative curvature of the spacetime has been realized early
\cite{Buchbinder:1989fz}, and is meanwhile reviewed in textbooks
\cite{Buchbinder:1992rb}. The phenomenon is typically studied at mean-field
level and occurs in many different fermionic models 
\cite{Inagaki:1993ya,Sachs:1993ss,Elizalde:1995kg,Kanemura:1995sx,Inagaki:1995bk,Inagaki:1996nb,Geyer:1996wg,Miele:1996rp,Inagaki:1997kz,Hashida:1999wb,Gorbar:2007kd,Hayashi:2008bm,Sasagawa:2012mn}. 
In \cite{Gorbar:1999wa}, the similarity to magnetic catalysis was realized in
terms of an effective dimensional reduction mechanism of the spectral
properties of the Dirac operator. This justifies the use of the terminology
``gravitational catalysis'' \cite{Ebert:2008pc}.

In fact, we find in the present work that the effective dimensional reduction
and the corresponding enhancement of the density of states in the IR is
directly related to the fixed point structure as identified below. From this
RG viewpoint, symmetry breaking arises as a consequence of the fact that the
coupling value required for criticality becomes arbitrarily small as a
function of the curvature (the catalyzer). Hence, any finite value of the
fermionic interactions ultimately becomes supercritical, typically driving the
system towards the ordered phase. 

We investigate this RG mechanism within the simple $3d$ Gross-Neveu
model in the present work. We concentrate on two different curved
backgrounds with constant negative curvature: a maximally symmetric
spacetime (Anti de Sitter) and a purely spatially curved case
(Lobachevsky plane). For both cases, mean-field studies are already
available, see \cite{Inagaki:1995bk} and
\cite{Inagaki:1996nb,Gorbar:2007kd}, respectively. Whereas the former
allows for an analytic treatment in terms of simple functions, the
latter is potentially relevant for curved layered condensed matter
systems. For instance, the exitonic or anti-ferromagnetic
instabilities in graphite and graphene have been associated with
quantum phase transitions falling into the $3d$ Gross-Neveu
universality class \cite{Khveshchenko:2001zz,Herbut:2006cs}. As
catalyzed symmetry breaking is manifestly driven by the long-range
modes, the RG analysis allows us to estimate the required curvature in
relation to the length scale of the sample.

This paper is organized as follows: in Sect. \ref{sec:GNmodel}, we
briefly introduce the model and its formulation in curved
space. Section \ref{sec:FRG} is devoted to an evaluation of the RG
flow in its simplest formulation. The manifestation of gravitational
catalysis is discussed in Sect. \ref{sec:GC}. We estimate the
influence of finite probe length on gravitational catalysis in
Sect. \ref{sec:pseudo} by means of a pseudo-critical
coupling. Conclusions are drawn in Sect. \ref{sec:conc}. Relevant
technical details are deferred to the appendices.

\section{Gross-Neveu model in curved space}
\label{sec:GNmodel}

We aim at investigating the $3d$ Gross-Neveu model \cite{Gross:1974jv} in curved
spacetime with metric $g_{\mu \nu}$ (Greek indices running from $0$ to $2$)
and signature $(-,+,+)$ using functional RG methods. The microscopic action functional
$S$ at some ultraviolet (UV) scale $\Lambda$ depends on the bare coupling constant
$\bar{\lambda}_{\Lambda}$, the $\Nf$ Grassmann-valued fields $\psi =
(\psi^{i})$ and the $\Nf$ conjugated fields $\bar{\psi} =
(\bar{\psi}^{i})$,
\begin{align}
 S[\bar{\psi}^{i}, \psi^{i}] ={}& \!\! \regint{x} \! \left[
 \sum\limits_{i = 1}^{\Nf} \bar{\psi}^{i} \slashed{\nabla} \psi^{i} +
 \frac{\bar{\lambda}_{\Lambda}}{2 \Nf} \left( \sum\limits_{i = 1}^{\Nf} \bar{\psi}^{i} \psi^{i} \right)^{\!\!\!2} \right]
 \notag \\
 ={}& \!\! \regint{x} \left[ \bar{\psi} \slashed{\nabla} \psi +
   \frac{\bar{\lambda}_{\Lambda}}{2 \Nf} ( \bar{\psi} \psi )^2 \right] \!
 \text{,}
\label{eq:GNaction}
\end{align}
where $\regint{x} = \int \!\! \mrd^d x \sqrt{-g}$ is a shorthand for the
integral over the $d$-dimensional  spacetime and $g = \det g_{\mu \nu}$ is
the determinant of the spacetime metric. The differential operator
$\slashed{\nabla} = \gamma^{\mu} \nabla_{\mu}$ is composed from a set of
$d_\gamma$-dimensional gamma matrices satisfying the Clifford algebra
\begin{align}
 \{ \gamma_{\mu} , \gamma_{\nu} \} = 2 g_{\mu \nu} \mrI \text{,}
\end{align}
where $\mrI$ is the identity in spinor space. In the present work, we
use the irreducible representation, such that $d_\gamma=2$ specifies
the dimension of the $\gamma$ matrices as well as the number of Dirac
components of the fermions. For our computations with fermions in
curved space, we have actually used the Weldon formalism
\cite{Weldon:2000fr} which can be viewed as a generalization of the
conventional vierbein formalism. However, a mapping to standard
vierbein formulas is straightforward. The covariant derivative in the
Weldon formalism reads
\begin{align}
 \nabla_{\mu} \psi^{i} = \partial_{\mu} \psi^{i} + \Gamma_{\mu} \psi^{i} \text{,}
\end{align}
accounting for covariance with respect to the spinor and spacetime
structure. Here, $\Gamma_{\mu}$ is the affine spin connection, implicitly
defined by
\begin{align}\label{eq:defspincon}
\begin{aligned}
 \rmi \quad{}& 0 = \nabla_{\mu} \gamma^{\nu} = \partial_{\mu} \gamma^{\nu} + \Gamma_{\mu \rho}^{\nu} \gamma^{\rho} + [ \Gamma_{\mu} , \gamma^{\nu} ],\\
 \rmii \quad{}& 0 = \tr \Gamma_{\mu} \text{,}
\end{aligned}
\end{align}
with $\Gamma^{\nu}_{\mu \rho}$ the Christoffel symbols.\footnote{More
  generally, the spin connection $\Gamma_{\mu}$ can additionally
  accommodate a $U(1)$ gauge field $\mcA_{\mu}$ \cite{Weldon:2000fr}. In
  this case, the right-hand side of $\rmii$ in (\ref{eq:defspincon})
  would be given by $\tr \Gamma_{\mu} = - \cplx d_{\gamma} q
  \mcA_{\mu}$, where $q$ is the charge under the $U(1)$ gauge
  group. In the present work, we ignore such a $U(1)$ gauge sector,
  setting $\mcA_{\mu}=0$.}  In the present work, we are interested in
a discrete ``chiral'' $\mathds{Z}_2$ symmetry, where the
nontrivial transformation is defined by \cite{Hofling:2002hj}
\begin{equation}
\psi(x) \to - \psi(-x), \quad \bar\psi(x) \to \bar\psi
(-x). \label{eq:trafo5}
\end{equation}
This symmetry acts simultaneously on all flavors. It can spontaneously
be broken by a chiral condensate $\langle \bar\psi\psi\rangle\neq 0$,
which for finite interactions goes along with a mass gap generation.
Incidentally, the $3d$ Gross-Neveu model actually has a much larger
continuous U($\Nf$) flavor symmetry also allowing for more complicated
breaking patterns \cite{Janssen:2012pq}.\footnote{In many $3d$
  condensed matter systems where the Gross-Neveu model is considered
  as an effective theory, the low-energy degrees of freedom can be
  arranged into $\Nffour$ 4-component Dirac spinors, corresponding to
  a reducible representation of the Dirac algebra. This reducible
  representation can be constructed from a suitable combination of
  2-component spinors such that $\Nf =2 \Nffour$ in terms of the
  counting of fermions of the present work, see, e.g.,
  \cite{Janssen:2012oca} for a review. Note that the Gross-Neveu
  interaction term considered in this work $\sim (\bar\psi\psi)^2$
  corresponds to $\sim (\bar\psi\gamma_{45}\psi)^2$ in the reducible
  4-component notation of \cite{Janssen:2012pq} (or to $\sim
    (\bar\psi\gamma_{35}\psi)^2$ in the notation of
    \cite{Scherer:2012nn,Gorbar:2007kd}) for even $\Nf$. The critical
  properties of the discrete chiral transition are, however, identical
  to a 4-component Gross-Neveu model with a $(\bar\psi\psi)^2$
  interaction as considered in \cite{Braun:2010tt}.} 

\section{Fermionic RG flows in curved space}
\label{sec:FRG}

In the following, we use the functional renormalization group to compute the
RG flow of the Gross-Neveu coupling as a function of the (negative)
curvature. We employ the Wetterich equation \cite{Wetterich:1992yh}, describing
the flow of a scale-dependent effective action functional $\Gamma_k$ as a
function of an IR regulator scale $k$,
\begin{align}
 \partial_{k} \Gamma_{k}[\bar{\psi}^{i} , \psi^{i}] = \frac{\cplx}{2} \STr \left[ \left( \Gamma_{k}^\ttm{(2)} + R_{k} \right)^{-1} \partial_{k} R_{k} \right] \! \text{.} \label{eq:Wetteq}
\end{align}
The effective average action $\Gamma_k$ is related to the standard
generating functional for 1PI correlation functions in the limit
$\Gamma=\Gamma_{k\to 0}$. Towards the UV cutoff $k\to\Lambda$, $\Gamma_k$
approaches the microscopic action. The regularization is encoded in the
regulator function $R_k$, see below.  For reviews of the functional RG adapted
to the present context, see
Refs.~\cite{Berges:2000ew,Aoki:2000wm,Pawlowski:2005xe,Gies:2006wv,
  Delamotte:2007pf,Kopietz:2010zz,Metzner:2011cw,Braun:2011pp}. 

In the present work, we evaluate the flow within a rather simple approximation
for the effective action. For this, we truncate the effective action to
\begin{align}
 \Gamma_{k}[\bar{\psi}^{i}, \psi^{i}] =  \regint{x} \left[ \bar{\psi} \slashed{\nabla} \psi + \frac{\bar{\lambda}_{k}}{2 \Nf} ( \bar{\psi} \psi )^2 \right]\text{,}
\end{align}
where the only scale dependence lies in the four fermion coupling
$\bar{\lambda}_{k}$. Furthermore, $\bar{\lambda}_k$ parametrically depends on
the curvature of the background manifold.  The IR regularization is ensured by a chirally symmetric regulator of the form
\begin{align}
 R_k(x,y) ={}& \! \begin{pmatrix}
\slashed{\nabla} r(\tau) & 0 \\
0 & \slashed{\nabla}^{\mrT} r(\tau^{\mrT})
\end{pmatrix} \! \mathbb{1}(x,y) \text{,}\\
\mathbb{1}(x,y) ={}& \!
 \begin{pmatrix}
	\frac{\delta(x,y)}{\sqrt{-g}} & 0 \\
	0 & \frac{\delta(y,x)^{\mrT}}{\sqrt{-g}}
 \end{pmatrix} \! \text{,} \quad \tau = - \frac{\slashed{\nabla}^2}{k^2} \text{,}
\end{align}
where the superscript $\mrT$ denotes transposition in Dirac space, and
$\delta(x,y)$ represents a spin-valued delta distribution, keeping track of
the spinor or conjugate-spinor transformation properties associated with the
spacetime arguments. In App.~\ref{App:beta_function}, we briefly summarize
our conventions.

For practical computations, we use a Callan-Symanzik type regulator, that
facilitates the use of proper time representations,
\begin{align}\label{eq:csregulator}
r(x) = \sqrt{\frac{1 + x}{x}} - 1.
\end{align}
Within our investigations we restrict ourselves to negative curvature, giving
rise to gravitational catalysis. It is intuitively clear, that positive
curvature (e.g., a sphere) generically suppresses IR modes and thus reduces
the density of states of low lying modes
\cite{Buchbinder:1989fz,Buchbinder:1992rb}. We also consider the system mainly
in the large-$\Nf$ limit. At finite $\Nf$, further pointlike fermionic
self-interactions are generated which correspond to operators with an explicit
curvature dependence.\footnote{A similar mechanism has been observed in
  \cite{Scherer:2012nn} for the case of a magnetic field, and the
  corresponding operators have been classified.} We expect no qualitative
modifications from these operators and hence ignore them in the
following. This approximation becomes exact in the limit $\Nf\to\infty$.

\begin{figure}
\begin{center}
\includegraphics[width=0.4\textwidth]{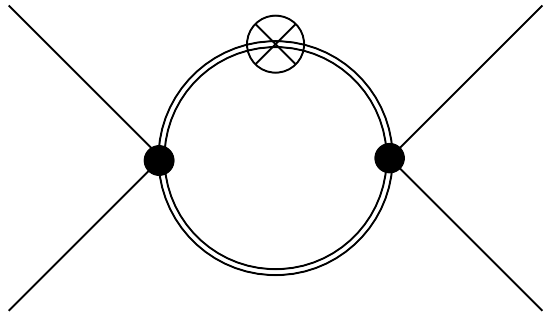} 
\end{center}
\caption{The diagram schematically exemplifies the fluctuation contributions
  to the flow of the Gross-Neveu coupling. The double lines represent the
  fermion propagator on the curved manifold. The full circles denote the
  RG-improved couplings, and the crossed circle marks the regulator
  insertion.}\label{fig:feynman_graph}
\end{figure}

It is straightforward (cf. App.~\ref{App:beta_function}) to calculate the flow
of the coupling as an implicit functional of the choice of the manifold,
which enters via the spectrum of the Dirac operator
\begin{align}\label{eq:betafunction1}
 \partial_{k} \bar{\lambda}_{k} = - \cplx \frac{2 \bar{\lambda}_{k}^2}{\Omega
   \Nf k^3} \STr \left[ \left( \mrI + \tau \right)^{-2}
   \frac{\delta(x,y)}{\sqrt{-g}} \right]. 
\end{align}
For this calculation, it suffices to project the flow onto constant fields
$\psi^{i}(x) \equiv \Psi^{i}$, with $\partial_\mu \Psi^i=0$. Here, $\Omega = \regint{x} 1$ denotes the
spacetime volume. The operator occurring in the trace is related to the square
of the regularized fermionic Green's function in curved spacetime. This has a
direct correspondence to a Feynman diagram representation, see
Fig. \ref{fig:feynman_graph}, as the flow in the present simple truncation is
driven by a single fermion bubble (and RG-improved resummations thereof). In
the following we distinguish between the cases of a maximally symmetric
spacetime in Sect. \ref{sec:maximally_symmetric} which can be treated fully
analytically, and a negatively curved space in Sect. \ref{sec:flat_time},
which is a more interesting case in view of two-dimensional condensed matter
systems.

\subsection{Maximally symmetric spacetime}
\label{sec:maximally_symmetric}

The use of a Callan-Symanzik regulator shape function,
cf. Eq. (\ref{eq:csregulator}), facilitates to rewrite the right-hand side of
Eq. (\ref{eq:betafunction1}) in terms of a simple proper time
representation. Other shape functions would still permit to use a proper time
representation but would lead to more intricate $k$ dependencies. The Laplace
transform of the operator of (\ref{eq:betafunction1}) reads
\begin{align}
 \left( \mrI + \tau \right)^{-2} \frac{\delta(x,y)}{\sqrt{-g}} = - k^4 \int\limits_{0}^{\infty} \! \mrd s \, s \euler^{- \cplx s k^2} \euler^{\cplx s \slashed{\nabla}^2} \frac{\delta(x,y)}{\sqrt{-g}} \text{,}
\end{align}
which upon insertion into (\ref{eq:betafunction1}) yields
\begin{align}
 \partial_{k} \bar{\lambda}_{k} = \cplx \frac{2 \bar{\lambda}_{k}^2 k}{\Omega
   \Nf} \int\limits_{0}^{\infty} \! \mrd s \, s \,\euler^{- \cplx s
   k^2} \STr \left[ \euler^{\cplx s \slashed{\nabla}^2}
   \frac{\delta(x,y)}{\sqrt{-g}} \right] \text{.}
\label{eq:lambdabar}
\end{align}

The expression inside the super trace is known as the heat kernel $K(x,y;s)=
\euler^{\cplx s \slashed{\nabla}^2}\delta(x,y)/\sqrt{-g}$ of the
(squared) Dirac operator. It satisfies
\begin{align}\label{eq:heatkernel}
\begin{aligned}
 \rmi \quad{}& \partial_{s} K = \cplx \slashed{\nabla}^2 K,\\
 \rmii \quad{}& \lim\limits_{s \ttm{\searrow} 0} K = \frac{\delta(x,y)}{\sqrt{-g}},
\end{aligned}
\end{align}
and was calculated in \cite{Camporesi:1992tm} for any maximally symmetric space,
with Euclidean signature. For our case the solution to this equation can be
obtained in an easier way with a special ansatz (cf. App.~\ref{App:heat_kernel}):
\begin{align}
 K = \frac{\euler^{\cplx \frac{d_{\mrG}^2}{4 s}}}{\cosh w} \left( \frac{w}{s \, \sinh w} + \cplx \frac{\abs{R}}{12 \cosh w} \right) \frac{\euler^{- \cplx \frac{\pi}{4}}}{(4 \pi)^{\frac{3}{2}} \sqrt{s}} U \text{,}
\end{align}
where $w^2 = \frac{\abs{R} d_{\mrG}^2}{24}$, and $d_{\mrG}(x,y)$ is
the geodesic distance between the points $x$ and $y$. The parallel
transporter $U$ (Wegner-Wilson line) is defined by
\begin{align}
 U(x,y) = \mrP \exp \left( - \int\limits_{0}^{1} \! \mrd t \, \frac{\mrd z^{\mu}(t)}{\mrd t} \Gamma_{\mu}\big(z(t)\big) \right) \! \text{,}
\end{align}
where $\mrP$ denotes the path ordering prescription and $z(t)$ is the geodesic between $x$ and $y$ with $z(t=0) = x$ and $z(t=1) = y$.

Now we are able to calculate the supertrace as
\begin{align}
 \STr \! \left[ \euler^{\cplx s \slashed{\nabla}^2} \frac{\delta(x,y)}{\sqrt{-g}} \right] \! = - \frac{2 \Omega \Nf}{(4 \pi)^{\frac{3}{2}} \sqrt{s}} \! \left( \frac{1}{s} + \cplx \frac{\abs{R}}{12} \right) \! \euler^{- \cplx \frac{\pi}{4}} \text{,}
\end{align}
and finally get
\begin{align}
 \partial_{k} \bar{\lambda}_{k} 
 ={}& - \frac{\bar{\lambda}_{k}^2}{2 \pi} \left( 1 + \frac{\abs{R}}{24 k^2} \right) \! \text{.}
\end{align}
In terms of the dimensionless coupling,
\begin{align}
 \lambda_{k} = k \bar{\lambda}_{k},
\end{align}
we obtain the beta function $\beta_{\lambda}$
\begin{equation}
 \beta_{\lambda} = k \partial_{k} \lambda_{k} = \lambda_{k} -
 \frac{\lambda_{k}^2}{2 \pi} \left( 1 + \frac{\abs{R}}{24 k^2}
 \right). \label{eq:betafunction_maximally_symmetric}
\end{equation}
This is an ordinary differential equation that parametrically depends on the
curvature. It can be solved by straightforward integration:
\begin{equation}
 \lambda_{k} = \frac{k}{\Lambda} \frac{\lambda_{\Lambda}}{1 - \frac{\lambda_{\Lambda} }{2 \pi} \left( 1 -
     \frac{k}{\Lambda} \right) \left( 1 + \frac{\abs{R}}{24 k \Lambda}
   \right) }. \label{eq:intlambda}
\end{equation}
The initial value is given by the dimensionless coupling $\lambda_{\Lambda}$
which in terms of the initial bare Gross-Neveu coupling reads
$\lambda_\Lambda=\Lambda\bar{\lambda}_{\Lambda}$.

\subsection{Negatively curved space}\label{sec:flat_time}

For the case of a manifold where the spatial part has a constant negative
curvature, we choose a special set of coordinates such that the metric can be
expressed as
\begin{align}
 (g_{\mu \nu}) =
\begin{pmatrix}
 -1 & \begin{matrix} 0 & 0 \end{matrix} \\
 \begin{matrix}
  0 \\
  0
 \end{matrix} & \!\!\!\! \text{\Large$\hat{g}_{i j}$}
\end{pmatrix} \text{,}
\end{align}
where $\hat{g}_{ij}$ (Latin indices running from $1$ to $2$) represents the
metric of a two dimensional maximally symmetric space and therefore only
depends on the spatial coordinates. Hence the gamma matrices are time
independent as well. The Christoffel symbols vanish for every time component
$\Gamma^{\rho}_{\mu 0} = \Gamma^{0}_{\mu \nu} = 0$ and also the curvature
tensor vanishes if any index is zero $R_{\mu \nu \rho 0} = 0$. From 
Eq.~(\ref{eq:defspincon}) $\rmi$, we infer that the spin connection
$\Gamma_{0}$ has to be proportional to $\mrI$, 
\begin{align}
 0 = \nabla_{0} \gamma^{\nu} = [\Gamma_{0} , \gamma^{\nu} ].
\end{align}
Moreover, From \Eqref{eq:defspincon} $\rmii$, we conclude that $\Gamma_{0}$
even has to vanish completely,
\begin{align}
 \tr \Gamma_{0} = 0 \text{.}
\end{align}
This implies that the operator $\slashed{\nabla}^2$ is separable into
\begin{align}
 \slashed{\nabla}^2 \psi^{i} ={}& - \partial_{0}^2 \psi^{i} + \vec{\slashed{\nabla}}^2 \psi^{i}\text{,} \quad \vec{\slashed{\nabla}} \psi^{i} = \gamma^k \nabla_k \psi^{i} \notag\\
 ={}& - \partial_{0}^2 \psi^{i} + \vec{\nabla}^2 \psi^{i} - \frac{R}{4} \psi^{i} \text{,}
\end{align}
where the curvature is only induced by the spatial components.

For a simpler calculation of the beta function, we perform a Wick rotation
$x^{0} \rightarrow - \cplx x^{0}$ and perform again a Laplace transformation
of the operator occurring in Eq. (\ref{eq:betafunction1}),
\begin{align}
 \left( \mrI + \tau \right)^{-2} \frac{\delta(x,y)}{\sqrt{-g}} = k^4 \int\limits_{0}^{\infty} \! \mrd s \, s \euler^{- s k^2} \euler^{s \slashed{\nabla}^2} \frac{\delta(x,y)}{\sqrt{-g}} \text{.}
\end{align}
We arrive at the Euclidean analogue of \Eqref{eq:lambdabar}
\begin{align}\label{eq:betafunction2}
 \partial_{k} \bar{\lambda}_{k} = \frac{2 \bar{\lambda}_{k}^2 k}{\Omega \Nf} \int\limits_{0}^{\infty} \! \mrd s \, s \euler^{- s k^2} \STr \left[ \euler^{s \slashed{\nabla}^2} \frac{\delta(x,y)}{\sqrt{-g}} \right] \text{.}
\end{align}
Here, we calculate the super trace again with the aid of the heat kernel using
that the differential operator $\slashed{\nabla}^2$ is separable and the delta
distribution factorizes in a time like and a spatial part
\begin{align}
 \euler^{s \slashed{\nabla}^2} \frac{\delta(x,y)}{\sqrt{- g}} = \euler^{s \partial_{0}^2} \delta(x^0 - y^0) \, \cdot \, \euler^{s \vec{\slashed{\nabla}}^2} \frac{\delta(\vec{x},\vec{y})}{\sqrt{{}\hat{g}}} \text{,}
\end{align}
where $\hat{g} = \det \hat{g}_{i j}$. The quantity $\vec{x}$ denotes the
spatial coordinates of $x$ and should be treated as a set of coordinates and
not as a vector. Both factors satisfy a heat-kernel equation and can be solved
analytically, cf. \cite{Camporesi:1992tm},
\begin{align}
 \euler^{s \partial_{0}^2} \delta(x^0 - y^0) &{}= \frac{\euler^{-\frac{(x^{0}-y^{0})^2}{4 s}}}{\sqrt{4 \pi s}} \text{,}\\
 \euler^{s \vec{\slashed{\nabla}}^2} \frac{\delta(\vec{x},\vec{y})}{\sqrt{\hat{g}}} ={}& \frac{2 \cosh^{-1} \! \frac{\hat{w}}{2}}{(4 \pi s)^{\frac{3}{2}} \sqrt{\abs{R}}} \!\! \int\limits_{\hat{w}}^{\infty} \! \mrd v \, \frac{v \euler^{- \frac{v^2}{2 s \abs{R}}} \cosh \frac{v}{2}}{\sqrt{\cosh v \! - \! \cosh \hat{w}}} \hat{U} \text{,}
\end{align}
where $\hat{w}^2 = \frac{\abs{R} \hat{d}_{\mrG}^2}{2}$, $\hat{U}$ is the parallel transporter for the spatial part and $\hat{d}_{\mrG}$ is the nonnegative spatial geodesic distance between $\vec{x}$ and $\vec{y}$ with $d_{\mrG}^{\,2}(x,y) = \hat{d}_{\mrG}^{\,2}(\vec{x}, \vec{y}) - (x^0 - y^0)^2$. Plugging these relations into Eq. (\ref{eq:betafunction2}) gives
\begin{align}
 \partial_k \bar{\lambda}_{k} ={}& - \frac{\bar{\lambda}_{k}^2}{2 \pi} \cdot \mfrI(\alpha_{k}) \text{,} \quad \alpha_{k} = \sqrt{\frac{\abs{R}}{2 k^2}},\\
 \mfrI(\alpha) ={}& \frac{\alpha}{2 \pi} \int\limits_{0}^{\infty} \! \mrd s \int\limits_{0}^{\infty} \! \mrd v \, \frac{\euler^{- s - \frac{v^2}{4 s}}}{s} v \coth \frac{\alpha v}{2} \text{.}
\end{align}
The $s$ integral is an integral representation of the modified Bessel
function of the second kind $K_{0}(v)$ \cite{Gradshteyn:2000}.
Therefore, we have
\begin{align}
 \mfrI(\alpha) = \frac{\alpha}{\pi} \int\limits_{0}^{\infty} \! \mrd v \, v K_{0}(v) \coth \frac{\alpha v}{2} \text{.}\label{eq:defJ}
\end{align}
With these results we are again able to derive the beta function for the
dimensionless coupling $\lambda_k=k \bar{\lambda}_k$,
\begin{equation}
 \beta_{\lambda} =k \partial_k \lambda_k= \lambda_{k} - \frac{\lambda_{k}^2}{2
   \pi} \cdot \mfrI (\alpha_{k}). \label{eq:betafunction_flat_time}
\end{equation}
The integration of this ordinary differential equation depending
parametrically on the curvature can be cast into an integral representation,
\begin{equation}
\lambda_{k} =\frac{k}{\Lambda} \frac{ \lambda_{\Lambda}} {1 -
  \frac{\lambda_{\Lambda}}{2 \pi} \alpha_{\Lambda} \int\limits_{\alpha_{\Lambda}}^{\alpha_{k}}
  \frac{\mfrI(\alpha)}{\alpha^2} \mrd \alpha }, \quad \alpha_k=
\sqrt{\frac{|R|}{2 k^2}}. \label{eq:intlambdaspat}
\end{equation}
This result is qualitatively similar to the maximally symmetric case of
\Eqref{eq:intlambda}.

\section{Gravitational catalysis}
\label{sec:GC}

\begin{figure}
\centering
  \psfrag{k}{$\lambda_k$}
  \psfrag{g}{}
  \psfrag{b}{\rotatebox[origin=c]{90}{$\beta_{\lambda}$}}
  \psfragnumbers
\includegraphics[width=0.48\textwidth]{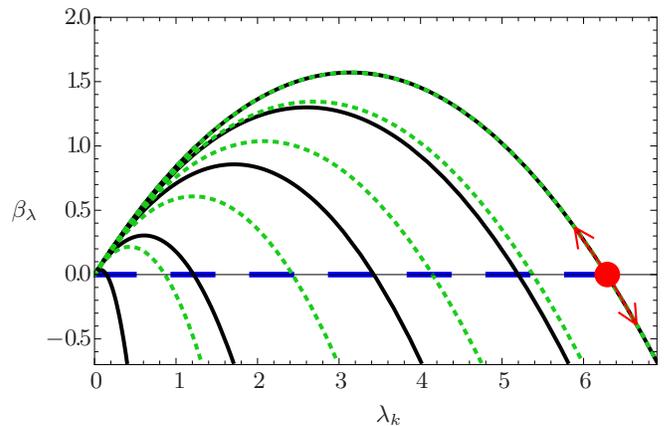} 
\caption{Plot of the RG $\beta_{\lambda}$ function of the coupling
  $\lambda_k$ for different values of the scale dependent negative
  curvature $\frac{\abs{R}}{k^2}$ (from top to bottom: $0; \, 5 ; \,
  20; \, 100 ; \, 1000 $). The black lines depict the
  $\beta_{\lambda}$ functions for the maximally symmetric spacetime
  (AdS), cf. Eq. (\ref{eq:betafunction_maximally_symmetric}); arrows
  indicate the flow towards the IR. The green dotted graphs show the
  flows for the case of purely spatial curvature (Lobachevsky plane),
  cf. Eq. (\ref{eq:betafunction_flat_time}). In addition to the
  Gau\ss{}ian fixed point, there exists a non Gau\ss{}ian fixed point
  (full red circle at $\lambda_{\text{cr}}=2\pi$ for the present
  regulator scheme), separating the symmetric phase for
  $\lambda_\Lambda<\lambda_{\text{cr}}$ from the broken phase for
  $\lambda_\Lambda>\lambda_{\text{cr}}$ for zero curvature.  For
  finite curvature, this critical point becomes scale-dependent and
  moves towards the Gau\ss{}ian fixed point for increasing scale
  dependent curvature, i.e., with decreasing IR scale for fixed
  curvature. In the case of vanishing curvature, the symmetry is
  preserved and no mass is generated for initial values
  $\lambda_\Lambda$ in the blue dashed
  region.}\label{fig:betafunction}
\end{figure}

Let us now analyze the consequences of the RG flows for the long-range
properties of the Gross-Neveu model.  For both background manifolds,
$\beta_{\lambda}$ considered as a function of $\lambda_{k}$ is a
parabola where the prefactor of the quadratic part is scale and
curvature dependent, see Fig. \ref{fig:betafunction}. For vanishing
curvature, the $\beta_{\lambda}$ function vanishes at the two fixed
points $\lambda_{k} = 0$ (Gau\ss{}ian) and $\lambda_{\ast}( R = 0 ) =
\lambda_{\mathrm{cr}} = 2 \pi$ which corresponds to the well-known
critical coupling of the Gross-Neveu model in flat space in this
regularization scheme \cite{Braun:2010tt,Braun:2011pp}. This critical
coupling separates the symmetric phase for $\lambda_{\Lambda} <
\lambda_{\mathrm{cr}}$ where the long range behavior is controlled by
the non-interacting Gau\ss{}ian fixed point from the chiral symmetry
broken phase for $\lambda_{\Lambda} > \lambda_{\mathrm{cr}}$. In the
latter case, $\lambda_{k}$ runs to large values towards the
infrared. In the present simple truncation, $\lambda_{k}$ in fact
diverges at a finite scale $k_{\text{SB}}$ signaling the transition
into the ordered regime. The scale $k_{\text{SB}}$ is thus
characteristic for the physical scales in the ordered phase. In
\cite{Jaeckel:2002rm} it has been shown that $k_{\text{SB}}$ actually
agrees with the value of the dynamically generated fermion mass
$m_{\text{f}}$ as obtained in mean-field approximation. Since we are
working in the large-$\Nf$ limit anyway, we will use this mean-field
identification in the following: $m_{\text{f}}=k_{\text{SB}}$.

The existence of the non-Gau\ss{}ian fixed point
$\lambda_{\mathrm{cr}}$ can be attributed to the competition between
the power-counting scaling (the linear coupling term in
$\beta_{\lambda}$) and the interaction terms $\sim \lambda_{k}^2$. In
our RG picture, the interaction terms are enhanced by negative
curvature as soon as the wavelength of the fluctuations becomes of the
order of the curvature scale. As a consequence, the interacting second
zero of the $\beta_\lambda$ function no longer is a true fixed point
but becomes scale dependent. This ``pseudo-critical coupling''
$\lambda_{\text{p}}=\lambda_{\ast} ( \abs{R} / k^2 )$ moves towards
the Gau\ss{}ian fixed point for decreasing scale $k$, see  Fig. \ref{fig:betafunction}.

Any finite initial coupling strength $\lambda_{\Lambda}$ will eventually
become larger than $\lambda_{\ast} ( \abs{R} / k^2 )$ for small RG scales $k
\rightarrow 0$. By this mechanism, the system is forced into the
symmetry-broken phase even at the weakest initial coupling. We observe this
mechanism in both cases of negative curvature, the maximally symmetric as well
as the purely spatial curvature case.
As we see in Fig. \ref{fig:betafunction} the influence of the curvature is
somewhat stronger in the maximally symmetric case. 

As discussed above, we calculate the symmetry breaking scale
  $k_{\text{SB}}$ by searching for a zero of the inverse coupling.  The
fermion mass $m_{\text{f}}$ corresponding to this scale where the RG flow
enters the symmetry-broken regime can thus be computed from the criterion
\begin{equation}
  \lambda_{k = m_{\text{f}}}^{-1} ( \abs{R} / m_{\text{f}}^2) = 0.\label{eq:critcond}
\end{equation}
Upon partial bosonization (Hubbard-Stratonovich transformation),
$\lambda_k^{-1}$ is related to the mass parameter of a composite bosonic
field. Hence the divergence of the fermionic self-interaction simply
corresponds to onset of the order parameter
\cite{Ellwanger:1994wy,Gies:2001nw,Braun:2011pp}.  Let us now analyze the two
different backgrounds under consideration in detail.

\subsection{Maximally symmetric spacetime}
\label{sec:maximally_symmetric2}

In the maximally symmetric case, the fermion mass defined by the
criterion \Eqref{eq:critcond} can be straightforwardly
computed from the running coupling \Eqref{eq:intlambda}, yielding
\begin{align}\label{eq:kc_maximally_symmetric}
\frac{m_{\text{f}}}{\Lambda} = \frac{1}{2} - \frac{\abs{R}}{48
  \Lambda^2} - \frac{\pi}{\lambda_{\Lambda}} + \sqrt{ \! \left(
  \frac{1}{2} - \frac{\abs{R}}{48 \Lambda^2} -
  \frac{\pi}{\lambda_{\Lambda}} \right)^{\!2} \! + \frac{\abs{R}}{24
    \Lambda^2}} \text{.}
\end{align}
Plots of this gravitationally catalyzed fermion mass are shown as a
function of the curvature as solid lines in
Fig.~\ref{fig:comparison_generated_mass}. Let us discuss this result in
various limits. 
In the zero-curvature limit $R=0$, we find $m_{\text{f}}=0$ for  $\lambda_{\Lambda}
\leq 2 \pi$. For super-critical couplings $\lambda_{\Lambda}>2\pi$, we
rediscover the standard mean-field result in $3d$,
\begin{equation}
m_{\text{f},0}\equiv m_{\text{f}}(R=0)= \Lambda \left(1 - \frac{2\pi}{\lambda_\Lambda}\right). 
\label{eq:meanres}
\end{equation}
This is in perfect agreement with the known behavior in flat spacetime.

Provided the fermion system is initially weakly
coupled, $\lambda_\Lambda\ll \lambda_{\text{cr}}=2\pi$, a leading
order expansion can be performed for any value of the curvature,
resulting in
\begin{equation}
m_{\text{f}} \simeq \frac{\Lambda}{1 +
  \frac{48\pi\Lambda}{ \vphantom{ \left[ \frac{A}{B} \right] } |R| \bar{\lambda}_\Lambda } }, \quad \text{for}\,\, \bar{\lambda}_\Lambda  \Lambda \ll 1,
\end{equation}
where we have reinserted the dimensionful initial coupling
$\bar{\lambda}_\Lambda=\lambda_\Lambda/\Lambda$. If we additionally
consider the limit of small curvature, we find a linear dependence
of the fermion mass on both the curvature as well as the coupling,
\begin{equation}
m_{\text{f}} \simeq \frac{1}{48\pi} |R| \bar{\lambda}_\Lambda.
\label{eq:weak}
\end{equation}
By contrast, in the limit of large curvature, $|R|/(48\pi \Lambda^2)
\gg \pi/\lambda_\Lambda \gg 1$, we find that $m_{\text{f}} \to
\Lambda$. In other words, large curvature induces immediate chiral
symmetry breaking, such that the induced mass becomes of the order of
the cutoff. Incidentally, this result is similar for the
large-coupling limit: for $\lambda_\Lambda\gg 2\pi$, we again find
that $m_{\text{f}} \to \Lambda$ to leading order independently of the
curvature.

The above results display explicit UV cutoff and regularization-scheme
dependencies. Since the $3d$ Gross-Neveu model is asymptotically safe
and thus non-perturbatively renormalizable \cite{Scherer:2012nn}, we
can remove the UV cutoff by keeping an IR obvservable fixed while
sending $\Lambda\to\infty$. This ``line of constant physics'' defines
a renormalized trajectory. This can most conveniently be done in the
super-critical regime where $\lambda_\Lambda>2\pi$ such that the
fermion mass in flat-space $m_{\text{f},0}$ of \Eqref{eq:meanres}
defines a natural IR renormalization point.\footnote{In the
  sub-critical regime, the model is quasi conformal and can be
  renormalized, e.g., by fixing the coupling $\lambda_k$ at a
  suitable renormalization point $k=\mu$ to a specific value.} In this
case, the generated fermion mass in the limit $\Lambda\to \infty$ can
be written as
\begin{eqnarray}
\frac{m_{\text{f}}}{m_{\text{f},0}} &=& \frac{1}{2} + \sqrt{\frac{1}{4} + \frac{|R|}{24 m_{\text{f},0}^2}} \nonumber\\
 &\simeq& 1+ \frac{|R|}{24 m_{\text{f},0}^2}, \label{eq:mfren}
\end{eqnarray}
where the first line holds for arbitrary curvature, and the
  second line represents a weak-curvature expansion being in perfect
  agreement with \cite{Inagaki:1995bk}.

We emphasize that the fermions acquire a mass $m_{\text{f}}>0$ for any
given $\lambda_{\Lambda}$ as long as the curvature is
nonvanishing. While we expect the tendency to drive the fermion system
towards the broken phase through gravitational catalysis to remain also
beyond our truncation, fluctuations of bosonic composites entering
beyond the large-$\Nf$ limit typically provide for an opposite
tendency. Hence, the status of gravitational catalysis beyond
mean-field remains an interesting question. Analogously, the effects
of beyond-mean-field fluctuations on magnetic catalysis are under
active current investigation \cite{Skokov:2011ib,Fukushima:2012xw}.

\subsection{Negatively curved space}\label{sec:flat_time2}

\begin{figure}
\centering
  \psfrag{k}{\large{\rotatebox[origin=c]{90}{$\frac{m_{\mrf}}{\Lambda}$}}}
  \psfrag{c}{}
  \psfrag{R}{\large{$\frac{\abs{R}}{\Lambda^2}$}}
  \psfragnumbers
\includegraphics[width=0.48\textwidth]{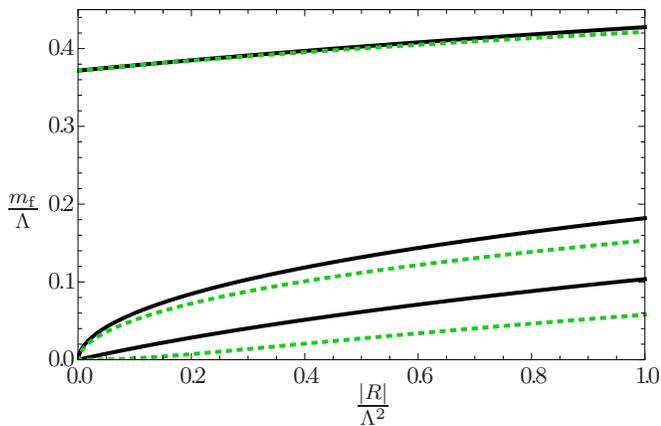}
\caption{Gravitationally catalyzed fermion masses (mean-field level) as a
  function of negative curvature in units of the UV cutoff. The solid black
  lines display the maximally symmetric case, whereas the purely
  spatially curved case is shown as green dotted lines. The sets of three different
  lines correspond to super-critical, critical, and sub-critical bare fermion
  couplings, $\lambda_{\Lambda} \simeq 1.6 \lambda_{\text{cr}},
  \lambda_{\text{cr}}, 0.8\lambda_{\text{cr}}$ from top to bottom
  ($\lambda_{\text{cr}}=2\pi$). As long as the background manifold is
  negatively curved, $|R|>0$, a finite fermion mass is
  generated.}\label{fig:comparison_generated_mass} 
\end{figure}

In the case of pure spatial curvature, the criterion
\Eqref{eq:critcond} cannot be resolved analytically, but we can give an
implicit equation for the induced fermion mass $m_{\text{f}}$
\begin{align}\label{eq:kc_flat_time}
  0 = \frac{2 \pi}{\lambda_{\Lambda}} - \alpha_{\Lambda}
  \int\limits_{\alpha_{\Lambda}}^{\alpha_{m_{\text{f}}}} \!
  \frac{\mfrI(\alpha)}{\alpha^2} \mrd \alpha \text{,}\quad
  \alpha_k=\sqrt{\frac{\abs{R}}{2 k^2}},
\end{align}
which can be solved numerically. Though the basic picture does not differ much
from the maximally symmetric case, there are some interesting differences. As can already be inferred from the beta function in
Fig.~\ref{fig:betafunction} (dotted lines), the curvature induced mass in the
spatially curved case is smaller than in the maximally symmetric case,
cf. Fig. \ref{fig:comparison_generated_mass}.
Several limits can be discussed in an analytic fashion, using the series
representation of $\mfrI(\alpha)$ developed in
App.~\ref{App:curvature_expansion}. Furthermore, we need the integral
$\mfrF(\alpha)$ defined by
\begin{align}\label{eq:def_mfrF}
\begin{aligned}
 \rmi \quad {}& \mfrF(\alpha) = \int \! \frac{\mfrI(\alpha)}{\alpha^2} \mrd \alpha,\\
 \rmii \quad {}& \lim\limits_{\alpha \rightarrow \infty} \left( \mfrF(\alpha) - \frac{\ln \alpha}{\pi} \right) = 0 \text{,}
\end{aligned}
\end{align}
where $\rmii$ fixes the constant of integration. The explicit calculation is
done in App.~\ref{App:curvature_expansion}. The defining equation for the
fermion mass in terms of $\mfrF$ is
\begin{align}
 \mfrF( \alpha_{m_{\text{f}}} ) = \mfrF(\alpha_{\Lambda}) + \frac{2 \pi}{\lambda_{\Lambda} \alpha_{\Lambda}} \text{.}\label{eq:calF}
\end{align}
This representation provides us with some interesting insight. First, we can
show that there exists a unique solution to this equation with $0 <
m_{\text{f}} < \Lambda$ for any given negative curvature $\abs{R}>0$ and
$\lambda_{\Lambda} > 0$. This can be seen in two steps. The uniqueness is
because the function $\mfrF(\alpha) \!\! \in \!\! (- \infty, \infty)$ for
$\alpha \!\! \in \!\!  (0,\infty)$ is bijective. This property holds, because
$\mfrI(\alpha)$ is positive and therefore $\mfrF$ is strictly monotonically
increasing. Owing to $\mfrI(\alpha)/\alpha^2 \rightarrow 1/\alpha^2$ for small
$\alpha \rightarrow 0$ and $\mfrI(\alpha)/\alpha^2 \rightarrow 1/(\pi \alpha)$
for large $\alpha \rightarrow \infty$, the function $\mfrF$ is not
bounded. Since by assumption $2 \pi/(\lambda_{\Lambda} \alpha_{\Lambda}) > 0$,
it follows from \Eqref{eq:calF} that $\mfrF(\alpha_{m_{\text{f}}}) >
\mfrF(\alpha_{\Lambda})$ has to hold, which implies that
$\alpha_{m_{\text{f}}} > \alpha_{\Lambda}$ because of the monotonic
behavior. This demonstrates that $0 < m_{\text{f}} < \Lambda$ as claimed
above.

Let us first check the flat spacetime limit $\abs{R} \rightarrow
0$. In complete agreement with \Eqref{eq:meanres}, we find again that
\begin{align}
  \frac{2 \pi}{\lambda_{\Lambda}} = \lim\limits_{\abs{R} \rightarrow 0} \left[
    \alpha_{\Lambda} \mfrF(\alpha_{m_{\text{f}}}) - \alpha_{\Lambda}
    \mfrF(\alpha_{\Lambda})\right] = 1 - \frac{m_{\text{f},0}}{\Lambda},
\end{align}
where we have used \Eqref{eq:mfrF_small_alpha}.

In the weak coupling regime, we expect the fermion mass to be small compared
to the curvature scale, cf. \Eqref{eq:weak}. Hence, we need $\mfrF(\alpha)$
for large argument as provided by \Eqref{eq:mfrF_big_alpha}, 
\begin{align}
 \mfrF(\alpha\gg 1) \simeq \frac{\ln \alpha}{\pi}.
\end{align}
Using this asymptotic behavior for $\mfrF(\alpha_{m_{\text{f}}})$, we infer
from \Eqref{eq:calF} that 
\begin{align}
m_{\text{f}}\simeq \sqrt{\frac{|R|}{2}} \exp \left[ -
    \frac{\pi}{\alpha_{\Lambda}} \left( \frac{2 \pi}{\lambda_{\Lambda}} +
      \alpha_{\Lambda} \mfrF(\alpha_{\Lambda}) \right) \right], \,\,
  \alpha_\Lambda= \sqrt{\frac{|R|}{2\Lambda^2}}.  
\end{align}
If in addition, the curvature is small, i.e., $\alpha_\Lambda \ll 1$,
we can use the corresponding expansion of \Eqref{eq:mfrF_small_alpha}
and replace $\alpha_{\Lambda} \mfrF(\alpha_{\Lambda}) \to
-1$. Taking differences arising from the Dirac representation
  into account, the exponential inverse-coupling dependence is in
  perfect agreement with the results of \cite{Gorbar:2007kd}.

Despite the overall similarities to the maximally symmetric case, we observe
that the weak-coupling and weak-curvature limit of the case with pure spatial
curvature shows some distinct differences. In particular, there is an
exponential non-analytic dependence of the fermion mass on the coupling as
well as on the curvature. 

This difference is reminiscent to magnetic catalysis in $d=2+1$ and
$d=3+1$ \cite{Gusynin:1994xp}, where the fermion gap is analytic in
$d=2+1$, but shows an essential singularity in the coupling in
$d=3+1$. Also in the present case, such a singularity shows up similar
to BCS gap formation, as a consequence of the effective dimensional
reduction of the fermionic fluctuation spectrum to $d\to 1+1$
\cite{Gorbar:1999wa}. Also in this respect, our functional RG
  picture is in perfect agreement with previous studies
  \cite{Gorbar:2007kd,Ebert:2008pc}.

\section{Pseudo-critical coupling and probe size}
\label{sec:pseudo}

At zero curvature, the $3d$ Gross-Neveu model exhibits a critical coupling
strength corresponding to a quantum critical point above which chiral symmetry
is broken at large length scales. This critical coupling manifests itself as a
non-Gau\ss ian fixed point of the RG flow. In the present regularization
scheme, we identified $\bar{\lambda}_{\text{cr}}=2\pi/\Lambda$, or
$\lambda_{\text{cr}}=2\pi$ in dimensionless conventions. As illustrated in
Fig.~\ref{fig:betafunction}, the fixed point strictly speaking no longer
exists at finite negative curvature. The nontrivial zero of the
$\beta_\lambda$ function becomes  scale dependent and eventually merges with
the Gau\ss ian fixed point in the deep IR for $k\to0$, such that only the
chirally broken branch of the $\beta_\lambda$ function remains. 

Let us call the nontrivial zero of $\beta_\lambda$ a pseudo-critical coupling
$\lambda_{\text{p}}$. We find
\begin{equation}
\lambda_{\text{p}}=\frac{\lambda_{\text{cr}}}{1+ \frac{|R|}{24k^2}}
\label{eq:lpMS}
\end{equation}
for the maximally symmetric case,
cf. \Eqref{eq:betafunction_maximally_symmetric}, and
\begin{equation}
\lambda_{\text{p}}=\frac{\lambda_{\text{cr}}}{\mfrI \! \left( \vphantom{\left[\frac{A}{B}\right]^{A}} \right. \!\! \sqrt{\frac{|R|}{2k^2}}  \left. \vphantom{\left[\frac{A}{B}\right]^{A}} \right)}
\label{eq:lpSC}
\end{equation}
for the spatially curved case, cf. \Eqref{eq:betafunction_flat_time}. Since
$\lambda_{\text{p}}= \lambda_{\text{p}}(k)$ is a monotonically decreasing
function of scale $k$, the coupling $\lambda_k$ can eventually exceed
$\lambda_{\text{p}}$ such that the system becomes critical and runs towards
the ordered phase. In this sense, 
\begin{equation}
\lambda_{k_{\text{c}}} = \lambda_{\text{p}}(k_{\text{c}})\label{eq:critc}
\end{equation}
can be viewed as a criticality condition \cite{Braun:2010qs}, defining a scale
$k_{\text{c}}$, where the system becomes critical. For lower scale, the system
is driven towards the symmetry broken regime which is ultimately entered at
$k_{\text{SB}}<k_{\text{c}}$ defined above. The value of $k_{\text{c}}$ depends on the
curvature as well as the initial coupling $\lambda_{\Lambda}$ (the latter is
considered as initially subcritical here and in the following).

The preceding discussion implicitly assumed that $k$ can run over all scales
down to $k=0$, such that the criticality condition \eqref{eq:critc} can
eventually always be satisfied. However, if, for instance, the system has a
finite volume characterized by a finite length scale $L$, also the fluctuation
momenta are restricted, typically leading to an IR cutoff $k_L=
\pi/L$.\footnote{Here, we tacitly assume that the boundary conditions are such
  that zero modes do not occur.} One may think of a finite probe length, such
as, e.g., the size of a layer of graphene. This finite probe length $L$ can
lead to a screening of the gravitationally catalyzed ordered regime if $k_L$
is larger than the would-be critical scale $k_{\text{c}}$. Hence,
$\lambda_{\text{p}}(k_L)$ can be viewed as a lower bound for the coupling
strength required for symmetry breaking in a real system of finite length. It
thus generalizes the critical coupling at infinite volume and zero curvature
to the situation of finite volume and finite curvature. 

It is instructive to study $\lambda_{\text{p}}$ as a function of the length
scale $L$ measured in units of a typical curvature length scale which we
define by $r=1/\sqrt{|R|}$. For finite systems, this gives an estimate for how strongly a
probe has to be curved in order to exhibit gravitational catalysis. In turn,
for a given curvature of the probe, $\lambda_{\text{p}}$ provides an estimate
for the initial coupling strength required for symmetry breaking in the finite
system. For instance, for the maximally symmetric case, we read off from
\Eqref{eq:lpMS} that
\begin{equation}
\lambda_{\text{p}}(k_L=\pi/L)=\frac{\lambda_{\text{cr}}}{1+ \frac{L^2}{24\pi^2
  r^2}}.
\label{eq:lpcurvMS}
\end{equation}
For the spatially curved case, we incidentally find the same result in
the limit of small curvature, i.e., $L/r\ll 1$, using the expansion
\eqref{eq:mfrI_small_alpha} which is accurate even for values of $L/r
\simeq \mathcal{O}(1)$. By contrast, the large curvature limit for the
spatially curved case is different, cf. \Eqref{eq:largecurve}:
\begin{equation}
\lambda_{\text{p}}(k_L=\pi/L)=\sqrt{2} \pi^2 \frac{r}{L} \lambda_{\text{cr}}, 
\quad \frac{r}{L}\ll 1.
\label{eq:lpcurvSC}
\end{equation}
From \Eqref{eq:lpcurvMS}, it is obvious that probe length to curvature
ratios up to $L/r\simeq\mathcal{O}(1)$ lead to pseudo-critical
couplings $\lambda_{\text{p}}$ which deviate from the zero-curvature
critical coupling of the Gross-Neveu model only below the 1\%
level. Significant deviations only occur for $L$ being an order of
magnitude larger than the curvature scale $r$. From the viewpoint of
curvature-deformed condensed matter systems, a ratio of
$L/r\simeq\mathcal{O}(1)$ appears to be ``large'' in the sense that --
loosely speaking -- a spatially curved $2d$ planar probe embedded in
$3d$ Euclidean space would rather look like a $3d$ object.

Another way to interpret these results is the following: consider a
finite-probe system with a subcritical bare coupling, $\lambda_\Lambda <
\lambda_{\text{cr}}$, thus being in the symmetric (e.g. semimetal) phase.  In
order to gravitationally catalyze a transition to a broken (gapped or
insulating) phase, the criticality condition \eqref{eq:critc} has to be met
for a sufficiently large $k_{\text{c}}>k_L$. In view of \Eqref{eq:lpcurvMS}
this requires comparatively strong curvature, i.e., a small curvature length scale $r$ compared with the probe length $L$.

\section{Conclusion and Outlook}
\label{sec:conc}

We have investigated the phenomenon of gravitational catalysis in the
$3d$ Gross-Neveu model on specific manifolds with constant negative
curvature. While the mechanism had already been studied frequently
with mean-field methods as well as from the viewpoint of the
fluctuation spectrum of the Dirac operator, we have added a new
renormalization group picture to the comprehensive understanding of
this phenomenon. The essence of this picture is that the critical
coupling of the fermionic system, corresponding to a quantum phase
transition in flat spacetime, is transmuted into a scale-dependent
pseudo-critical coupling that flows to zero as a consequence of
long-wavelength fluctuations (compared to the curvature scale). In
this manner, the infinite-volume fermion system becomes critical for
any arbitrarily weak coupling. 

We have identified the RG (pseudo-) fixed point mechanism for two
example manifolds of constant negative curvature: a maximally
symmetric spacetime (Anti de Sitter) and a purely spatially curved
case (Lobachevsky plane). Both manifolds support the mechanism
of gravitational catalysis, but exhibit a rather different behavior as
far as the dependence of chiral symmetry breaking on the coupling and
the curvature are concerned. The maximally symmetric case shows a
linear dependence (to leading order) on both quantities which makes
clear that the phenomenon is essentially perturbative, being
reminiscent of the ``quantum anomaly'' for fermions in a magnetic
field \cite{Gusynin:2005pk}. By contrast, order parameters indicating
the symmetry broken state such as the induced fermion mass exhibit an
essential singularity in both the coupling and the curvature for the
case of the purely spatially curved case. This is in many respects
similar to BCS-type gap formation. Again, our renormalization group
picture goes hand in hand here with properties of the fermionic
fluctuation spectrum, as analyzed in \cite{Gorbar:1999wa}.

As a benefit, the functional renormalization group also gives a
simple access to systems of finite extent by identifying the RG
infrared cutoff with an inverse length scale. In this manner, we can
estimate the fate of gravitational catalysis in finite systems. In
fact, the phenomenon only occurs, as long as the curvature radius is
sufficiently small compared to the systems length scale. We have been
able to phrase this statement quantitatively by introducing a
pseudo-critical coupling. Thinking in terms of curved layered
condensed matter systems, rather large curvatures are needed compared
to a realizable probe length in order to drive a sub-critical system
into a phase dominated by gravitational catalysis. 

We would like to emphasize that an immediate application of our results to
condensed matter systems would only be possible for reparametrization
invariant systems such as fluid membranes \cite{Nelson:1989xb},
curvature effects of which can be mapped onto the language of
Riemannian geometry. For tethered membranes or general lattice
systems, further phenomena connected to extrinsic curvature or
curvature related defects can be become relevant. In this context, it
is interesting to note that an external strain exerted on a graphene
sheet in flat space induces a pseudo-magnetic field \cite{Guinea:2009vd},
that may also support (pseudo-)magnetic catalysis. For such systems,
we hence expect an interesting interplay between these various effects
if we expose them to negative curvature inducing strain. 

Finally, gravitational catalysis may become relevant in the context of
asymptotically safe quantum gravity \cite{Weinberg:1980gg}. In
conjunction with fermionic degrees of freedom \cite{Dona:2012am}, the
UV fixed point determining the shape of the universe at highest
energies might go along with a negative (though scale-dependent)
curvature \cite{Lauscher:2005qz}. Whether or not gravitational
catalysis in connection with gravitionally modified critical fermion
interactions \cite{Eichhorn:2011pc} could become active and impose
constraints on the matter content of the universe then is a highly
involved question that deserves to be investigated in greater depth.

\acknowledgments 

The authors thank Astrid Eichhorn, Lukas Janssen, Daniel Scherer, Ren\'{e} Sondenheimer, and Andreas Wipf for valuable discussions and acknowledge support by
the DFG under grants Gi~328/5-2 (Heisenberg program), GRK1523 and FOR 723.


\appendix

\section{Derivation of the beta function}\label{App:beta_function}

Let us first introduce our conventions. In the Weldon formalism
\cite{Weldon:2000fr}, the Dirac conjugate spinor $\bar{\psi}$ is
related to the hermitean conjugate of the spinor $\psi$ via the
spin metric $h$
\begin{align}
 \bar{\psi} = \psi^{\dagger} h,
\end{align}
which is implicitly defined by
\begin{align}
\begin{aligned}
 \rmi \quad{}& h^{\dagger} = - h \text{,}\\
 \rmii \quad{}& \gamma_{\mu}^{\dagger} = - h \gamma_{\mu} h^{-1} \text{,}\\
 \rmiii \quad{}& \nabla_{\mu} h = \partial_{\mu} h - h \Gamma_{\mu} - \Gamma_{\mu}^{\dagger} h = 0 \text{.}
\label{eq:A2}
\end{aligned}
\end{align}
This gives rise to the following properties of the building blocks of
the Gross-Neveu action:
\begin{align}
\begin{aligned}
 \rmi \quad{}& \nabla_{\mu} \bar{\psi} = \partial_{\mu} \bar{\psi} - \bar{\psi} \Gamma_{\mu} \text{,}\\
 \rmii \quad{}& (\bar{\psi} \psi)^{\ast} = \bar{\psi} \psi \text{,}\\
 \rmiii \quad{}& \regint{x} (\bar{\psi} \slashed{\nabla} \psi)^{\ast} = \regint{x} \bar{\psi} \slashed{\nabla} \psi \text{.}
\end{aligned}
\end{align}
Choosing the spin metric $h$ to be anti-hermitean in \Eqref{eq:A2} (i) is
convenient for our metric and Clifford algebra conventions, but
differs from \cite{Weldon:2000fr}. Also, the absence of any imaginary
factor of ``$\cplx$'' in the fermion kinetic term is due to these
conventions. 

The flow equation \eqref{eq:Wetteq} uses a rather condensed
notation. More concretely, we work in field space parameterized by the
collective fields
\begin{equation}
 \phi ={} \begin{pmatrix}
         \psi \\ \bar{\psi}^{\mrT}
        \end{pmatrix} \text{,} \quad
 \bar{\phi} = \begin{pmatrix}
               \bar{\psi} \\ \psi^{\mrT}
              \end{pmatrix} \text{,}
\label{eq:fieldspace}
\end{equation}
representing Grassmann-valued functions on the manifold, reminiscent
to Nambu-Gorkov spinors. For instance, the classical action amended by
the regulator term which is used for deriving the flow equation from
the functional integral
(cf. \cite{Berges:2000ew,Aoki:2000wm,Pawlowski:2005xe,Gies:2006wv,
  Delamotte:2007pf,Kopietz:2010zz,Metzner:2011cw,Braun:2011pp}), reads
in these conventions
\begin{align}
 S_{k}[\phi] ={}& S[\phi] + \frac{1}{2} \regint{x} \!\!\! \regint{y} \bar{\phi}(x) R_{k}(x,y) \phi(y).
\end{align}
These conventions differ slightly from those commonly used, see, e.g.,
\cite{Gies:2001nw}, but turn out to be advantageous for coordinate
space computations on curved manifolds. For instance, the
representation of the unit element in field space becomes rather
intuitive,
\begin{align}
 \mathbb{1}(x,y) = \phi(x) \frac{\overleftarrow{\delta}}{\delta \phi(y)} = \frac{\overrightarrow \delta}{\delta \bar{\phi}(x)} \phi(y)
 = \! \begin{pmatrix}
    \frac{\delta(x,y)}{\sqrt{-g}} & 0 \\ 0 & \frac{\delta(y,x)^{\mrT}}{\sqrt{-g}}
   \end{pmatrix} \! \text{.}
\end{align}
By $\delta(x,y)$, we denote the
spin-valued delta distribution, which fulfills
\begin{align}
\begin{aligned}
 \rmi \quad{}& \psi(x) = \regint{y} \frac{\delta(x,y)}{\sqrt{-g}} \psi(y),\\
 \rmii \quad{}& \bar{\psi}(x) = \regint{y} \bar{\psi}(y) \frac{\delta(y,x)}{\sqrt{-g}}\text{.}
\end{aligned}
\end{align}
Therefore,
\begin{align}
 \nabla_{\mu}^{(x)} \frac{\delta(x,y)}{\sqrt{-g}} = \partial_{\mu}^{(x)} \frac{\delta(x,y)}{\sqrt{-g}} + \Gamma_{\mu} \frac{\delta(x,y)}{\sqrt{-g}}
\end{align}
has to hold. With this notation, we indicate that $\delta(x,y)$ transforms as
a spinor in $x$ and a Dirac-conjugated spinor in $y$, i.e. $\delta(x,y)$ can
be interpreted as $\delta(x,y) = \delta(x-y) U(x,y)$, with the standard scalar
delta distribution $\delta(x-y)$. We remark that there is a difference between
$\delta(x,y)$ and $\delta(y,x)^{\mrT}$ in the spinor structure, since
\begin{align}
 \nabla_{\mu}^{\mrT (x)} \frac{\delta(y,x)^{\mrT}}{\sqrt{-g}} ={}& \partial_{\mu}^{(x)} \frac{\delta(y,x)^{\mrT}}{\sqrt{-g}} - \Gamma_{\mu}^{\mrT} \frac{\delta(y,x)^{\mrT}}{\sqrt{-g}}\\
 ={}& \left( \nabla_{\mu}^{(x)} \frac{\delta(y,x)}{\sqrt{-g}} \right)^{\mrT} \text{.}
\end{align}
For the evaluation of the flow equation \Eqref{eq:Wetteq}, we proceed in a
standard fashion. We decompose
\begin{align}
 \Gamma_{k}^\ttm{(2)} + R_{k} = \mcF_{k} + \mcP_{k}
\end{align}
into a field-dependent part $\mcF_{k}$ and a field-independent part $\mcP_{k}$ 
in order to expand the Wetterich equation in powers of the fields,
\begin{align}\label{eq:rewetterich}
 \partial_{k} \Gamma_{k} = \frac{\cplx}{2} \sum\limits_{n = 0}^{\infty} (-1)^{n} \STr \left[ (\mcP_{k}^{-1} \partial_{k} R_{k}) (\mcP_{k}^{-1} \mcF_{k})^{n} \right] \text{.}
\end{align}
Since
\begin{align}
 \mcF_{k} ={}& \! - \frac{\bar{\lambda}_{k}}{\Nf} \!
 \begin{pmatrix}
	- [(\bar{\Psi} \Psi) \mrI + \Psi \bar{\Psi}] & \Psi \Psi^{\mrT} \\
	\bar{\Psi}^{\mrT} \bar{\Psi} & {} \!\!\! [(\bar{\Psi} \Psi) \mrI + \Psi \bar{\Psi}]^{\mrT} \!{}
 \end{pmatrix} \! \mathbb{1},\\
 \mcP_{k} ={}& \cplx k \!
 \begin{pmatrix}
	\sqrt{\tau}\big( \mrI + r(\tau) \big) & 0 \\
	0 & \sqrt{\tau^{\mrT}} \big( \mrI + r(\tau^{\mrT}) \big)
 \end{pmatrix} \! \mathbb{1},
\end{align}
we observe that only the term $n = 2$ in \Eqref{eq:rewetterich} can contribute
to the flow of $\bar{\lambda}_{k}$. We need
\begin{align}
 \mcP_{k}^{-1} \partial_{k} R_{k} = \frac{2}{k} \!
 \begin{pmatrix}
	\frac{\tau r'(\tau)}{\mrI + r(\tau)} & 0 \\
	0 & \frac{\tau^{\mrT} r'(\tau^{\mrT})}{\mrI + r(\tau^{\mrT})}
 \end{pmatrix} \! \mathbb{1}
\end{align}
with $r'(x) = \frac{\mrd}{\mrd x} r(x)$, and therefore only the diagonal of
$(\mcP_{k}^{-1} \mcF_{k})^2$ is required. In the limit $\Nf \rightarrow
\infty$, only the following terms remain:
\begin{align}
 &\left[(\mcP_{k}^{-1} \mcF_{k})^2\right]_{11}(x,y) = - \frac{\frac{\bar{\lambda}_{k}^2}{\Nf^{2} k^{2}} (\bar{\Psi} \Psi)^2}{\tau \big( \mrI + r(\tau) \big)^{2}} \, \frac{\delta(x,y)}{\sqrt{-g}},\\
 &\left[(\mcP_{k}^{-1} \mcF_{k})^2\right]_{22}(x,y) = \! \left[(\mcP_{k}^{-1} \mcF_{k})^2\right]^{\mrT}_{11}(y,x)\text{.}
\end{align}
The LHS of \Eqref{eq:rewetterich} boils down to
\begin{align}
 \partial_{k} \Gamma_{k} 
= \frac{\partial_{k} \bar{\lambda}_{k}}{2 \Nf} (\bar{\Psi} \Psi)^2 \Omega,
\end{align}
and the RHS yields
\begin{align}
 \frac{\cplx}{2} & \STr \! \left[ (\mcP_{k}^{-1} \partial_{k} R_{k}) (\mcP_{k}^{-1} \mcF_{k})^2 \right] \notag \\
 &{}= \cplx \frac{2 \bar{\lambda}_{k}^2}{\Nf^2 k^3} (\bar{\Psi} \Psi)^2 \STr \! \left[ \frac{ r'(\tau) }{ \big( \mrI + r(\tau) \big)^{3}} \frac{\delta(x,y)}{\sqrt{-g}} \right].
\end{align}
Inserting the Callan-Symanzik regulator \eqref{eq:csregulator}, we end up with
\Eqref{eq:betafunction1} of the main text.

\section{Derivation of the heat kernel}\label{App:heat_kernel}

Following \cite{Camporesi:1992tm}, we choose for the heat kernel satisfying
\Eqref{eq:heatkernel} the ansatz
\begin{align}
 K(x,y;s) = f\big( d_{\mrG}(x,y),s \big) \cdot U(x,y)
\end{align}
where $f$ is a scalar function of $d_{\mrG}$ the geodesic distance and the proper time $s$. Plugging this into equation (\ref{eq:heatkernel}) using $A = \sqrt{\frac{\abs{R}}{6}} \, \coth(2 w)$, $B = \sqrt{\frac{\abs{R}}{96}} \, \tanh w$ and
\begin{align}
\begin{aligned}
 \rmi \quad{}& n_{\mu} n^{\mu} = 1 \text{,} \quad n_{\mu} = \partial_{\mu} d_{\mrG},\\
 \rmii \quad{}& \partial_{\mu} n_{\nu} - \Gamma^{\rho}_{\mu \nu} n_{\rho} = A (g_{\mu \nu} - n_{\mu} n_{\nu}),\\
 \rmiii \quad {}& \nabla_{\mu} U = B [\gamma_{\mu} , \gamma_{\nu}] n^{\nu} U \text{,}
\end{aligned}
\end{align}
cf. \cite{Camporesi:1992tm}, we get
\begin{align}
 0 = \left( \frac{\abs{R}}{4} - 8 B^2 \right) f + 2 A f' + f'' + \cplx \dot{f}
\end{align}
with $f' = \partial_{d_{\mrG}} f$, $\dot{f} = \partial_{s} f$. Because of the boundary condition for $K$ and the regularity of $U(x,y=x) = \mrI$ we observe that
\begin{align}
 \lim\limits_{s \ttm{\searrow} 0} f = \frac{\delta(x-y)}{\sqrt{-g}} \text{,}
\end{align}
has to hold, with $\delta(x-y)$ representing the standard delta
distribution. One suitable representation for the delta distribution in 3$d$
maximally symmetric space with negative curvature in the limit $s \searrow 0$
is
\begin{align}
 \frac{\delta(x-y)}{\sqrt{-g}} = \frac{\euler^{- \cplx \frac{\pi}{4}}}{(4 \pi s)^{\frac{3}{2}}} \exp \left( \cplx \frac{d_{\mrG}^2}{4 s} \right).
\end{align}
Next, we factorize $f$ into the delta part and an auxiliary function $p$ of
$d_{\mrG}$ and $s$,
\begin{align}
 f(d_{\mrG},s) = \frac{p(d_{\mrG},s)}{\sqrt{s}} \frac{\euler^{- \cplx \frac{\pi}{4}}}{(4 \pi)^{\frac{3}{2}}} \exp \left( \cplx \frac{d_{\mrG}^2}{4 s} \right).
\end{align}
Expanding $p$ in powers of $\frac{1}{s}$ and using the boundary conditions,
\begin{align}
 p = \frac{1}{s} p_{1}(d_{\mrG}) + \cplx p_{0}(d_{\mrG}),
\end{align}
leads to
\begin{align}
\begin{aligned}
 \rmi \,\, {}& 0 = - \! \left( \frac{1}{d_{\mrG}} - A \right) \! p_{1} + p_{1}', \\
 \rmii \,\, {}& 0 = d_{\mrG} A p_{0} \! + \! d_{\mrG} p_{0}' \! + \! \! \left( \! 8 B^2 \! - \! \frac{\abs{R}}{4} \! \right) \! p_{1} \! - \! 2A p_{1}' \! - \! p_{1}'',\\
 \rmiii \,\, {}& 0 = \! \left( 8 B^2 - \frac{\abs{R}}{4} \right) \! p_{0} - 2 A p_{0}' - p_{0}'',
\end{aligned}
\end{align}
with the boundary condition $p_{1}(0) = 1$. From equation $\rmi$ and the
boundary condition, we see
\begin{align}
 p_{1} = \frac{2 w}{\sinh(2 w)} \text{.}
\end{align}
Plugging this into $\rmii$ gives
\begin{align}
 p_{0} = \frac{\abs{R}}{12 \cosh^2 w},
\end{align}
where we have eliminated the constant of integration using equation
$\rmiii$. From this, we finally get
\begin{align}
 K = \frac{\euler^{\cplx \frac{d_{\mrG}^2}{4 s}}}{\cosh w} \left( \frac{w}{s \, \sinh w} + \cplx \frac{\abs{R}}{12 \cosh w} \right) \frac{\euler^{- \cplx \frac{\pi}{4}}}{(4 \pi)^{\frac{3}{2}} \sqrt{s}} U \text{.}
\end{align}

\section{Curvature expansion}\label{App:curvature_expansion}

For the detailed analysis of the spatially curved case, the integral
representation of the running coupling \Eqref{eq:intlambdaspat} can be
studied in various limits analytically.  More specifically,
$\mfrI(\alpha)$ as defined in \Eqref{eq:defJ} can be expanded for
small and for large values of $\alpha$.  Let us first consider an
expansion of this function about $\alpha = 0$, starting with an
expansion of the integrand,
\begin{align}
 \alpha v \coth \frac{\alpha v}{2} \simeq 2 + \frac{1}{6} (\alpha v)^2 - \frac{1}{360} (\alpha v)^4 + \mcO\big((\alpha v)^6\big).
\end{align}
Using the standard integral
\begin{align}\label{eq:bessel_integral}
 \int\limits_{0}^{\infty} \! \mrd v v^{2 k} K_{0}(v) \! = \! 2^{2 k-1} \Gamma \! \left( \frac{2k+1}{2} \right)^{\!2} \!\!
 = \! \frac{\pi}{2^{2k+1}} \! \left( \! \frac{(2k)!}{k!} \right)^{\!2} \!\! \text{,}
\end{align}
the small $\alpha$ expansion of $\mfrI(\alpha)$ can be computed to any
order. To order $\alpha^4$, we find
\begin{align}\label{eq:mfrI_small_alpha}
 \mfrI(\alpha) \simeq 
 1 + \frac{\alpha^2}{12} - \frac{\alpha^4}{80} \text{.}
\end{align}
Due to the factorial growth of the coefficients,
cf. Eq. (\ref{eq:bessel_integral}), the expansion is an asymptotic series. By
comparison with the numerical result, the accuracy of
Eq. (\ref{eq:mfrI_small_alpha}) turns out to be above 99\% up to $\alpha
\simeq 1$.

A similar approximation can be done for large $\alpha$ by expanding
\begin{align}
 \coth \frac{\alpha v}{2} = \frac{1 + \euler^{- \alpha v}}{1 - \euler^{- \alpha v}} = 1 + 2 \sum\limits_{n = 1}^{\infty} \euler^{- n \alpha v} \text{,}
\end{align}
which holds for any $\alpha v > 0$. Next, we employ
\begin{align}
 \int\limits_{0}^{\infty}  \mrd v \, v K_{0}(v) \euler^{- n \alpha v} = \frac{- 1}{(n \alpha)^2 - 1} + \frac{n \alpha \arcosh (n \alpha)}{\big( (n \alpha )^2 - 1 \big)^{\frac{3}{2}}},
\end{align}
which can be used for $n \alpha > -1$. For $-1 <n \alpha < 1$, an analytic
continuation into the complex is implicitly understood, leading to a
replacement of the term $\arcosh( n \alpha ) \big( (n \alpha)^2 - 1
\big)^{-3/2}$ by $[- \arccos(n \alpha)] \big( 1 - (n \alpha)^2
\big)^{-3/2}$ here and in the following. This
leads to the convergent series
\begin{align}\label{eq:mfrI_sum_rep}
 \mfrI(\alpha) = \frac{\alpha}{\pi} + \frac{2}{\pi} \sum\limits_{n=1}^{\infty} \left[ \frac{- \alpha}{(n \alpha)^2 - 1} + \frac{n \alpha^2 \arcosh(n \alpha)}{\big( (n \alpha)^2 - 1 \big)^{\frac{3}{2}}} \right]
\end{align}
for any $\alpha \geq 0$.
Neglecting orders higher than $\frac{1}{\alpha}$ and using $\arcosh ( n \alpha ) \rightarrow \ln (2 n \alpha)$ for $n \alpha \rightarrow \infty$, we arrive at
\begin{align}
 \mfrI(\alpha) \simeq
 \frac{\alpha}{\pi} + \frac{\pi}{3} \frac{\ln \alpha}{\alpha} + \frac{\pi}{3}
 \left( 1 + \gamma - \ln \frac{A^{12}}{\pi} \right) \frac{1}{\alpha},
\label{eq:largecurve}
\end{align}
where $\gamma \approx 0.577$ is the Euler-Mascheroni constant and $A \approx
1.282$ is the Glaisher-Kinkelin constant. The accuracy of this result is above
99\% for $\alpha > 5$.

With the series (\ref{eq:mfrI_sum_rep}) it is possible to find a series representation for the required integral (\ref{eq:def_mfrF})
\begin{align}
 \mfrF(\alpha) = \frac{\ln \alpha}{\pi} + \frac{2}{\pi} \sum\limits_{n=1}^{\infty} \left[ \ln(2 n \alpha) - \frac{n \alpha \arcosh(n \alpha)}{\sqrt{(n \alpha)^2 - 1}} \right] \! \text{.}
\end{align}
This series is convergent for any $\alpha > 0$. The expansions for small and
large $\alpha$ are
\begin{align}
 \alpha \rightarrow 0: \quad{}& \mfrF(\alpha) \simeq - \frac{1}{\alpha} + c + \frac{\alpha}{12} - \frac{\alpha^3}{240}, \label{eq:mfrF_small_alpha}\\
 \alpha \rightarrow \infty: \quad{}& \mfrF(\alpha) \simeq \frac{\ln \alpha}{\pi} - \frac{\pi}{6} \frac{\ln \alpha}{\alpha^2}, \label{eq:mfrF_big_alpha}
\end{align}
where $c$ is defined as the limit
\begin{align}
 c = \lim\limits_{\alpha \rightarrow 0} \left( \mfrF(\alpha) + \frac{1}{\alpha} \right) \approx 0.364 \text{.}
\end{align}

\end{document}